# Automated grading and staging of ovarian cancer using deep learning on the transmission optical microscopy bright-field images of thin biopsy tissue samples

Ashmit K Mishra, Mousa Alrubayan and Prabhakar Pradhan

Department of Physics and Astronomy, Mississippi State University, MS 39762

**Abstract:** Ovarian cancer remains a challenging malignancy to diagnose and manage, with prognosis heavily dependent on the stage at detection. Accurate grading and staging, primarily based on histopathological examination of biopsy tissue samples, are crucial for treatment planning and predicting outcomes. However, this manual process is time-consuming and subject to inter-observer variability among pathologists. The increasing volume of digital histopathology slides necessitates the development of robust, automated methods to assist in this critical diagnostic step for ovarian cancer. (Methods) This study presents a deep learning framework for the automated prediction of ovarian cancer stage (classified into five categories: 0, I, II, III, IV) using routine histopathological images. We employed a transfer learning approach, fine-tuning a ResNet-101 convolutional neural network pre-trained on ImageNet. The training process incorporated comprehensive data augmentation, weighted random sampling, and class weighting to address dataset characteristics. Hyperparameter optimization for learning rate, dropout rate, and weight decay was performed using a genetic algorithm to enhance model performance and generalization. (Results) Evaluated on an independent test set of ovarian thin tissue brightfield images, the developed model achieved a high overall classification accuracy of 97.62%.

## 1. Introduction

Ovarian cancer is the most lethal gynecological malignancy, mainly due to its tendency to be diagnosed at advanced stages [1]. Accurate staging is the most important prognostic factor and dictates the primary treatment choice [2]. The surgical-pathological staging of ovarian cancer typically follows the International Federation of Gynecology and Obstetrics (FIGO) system, which assigns stages (I, II, III, IV) based on the extent of tumor spread within the pelvis and to distant sites. It often involves pathological assessment of tumor grade based on microscopic features [3,4]. Precise histopathological evaluation of the primary tumor and metastatic sites is therefore fundamental to the accurate staging of ovarian cancer [5].

Examining ovarian tissue slides for grading and staging is highly skilled but can be subjective, leading to inter-observer variability among pathologists, particularly in subtle or complex cases [6]. Moreover, the increasing volume of surgical pathology specimens places significant demands on pathology departments, contributing to diagnostic turnaround times. The advent of digital pathology, involving the digitization of glass slides into whole-slide images (WSIs), facilitates computational analysis and offers opportunities to develop automated tools to assist pathologists [7].

Artificial intelligence (AI), and specifically deep learning (DL) with Convolutional Neural Networks (CNNs), has shown immense promise in analyzing medical images, including complex histopathology slides or brightfield images [8,9]. Applications in computational pathology range from detecting specific features like mitotic figures and tumor budding to classifying different tumor types and predicting molecular subtypes [10]. While automated grading of specific cancer types (like ovarian cancer) using AI has seen considerable research, automated staging directly from optical transmission brightfield image features remains less explored, particularly for ovarian cancer [11]. Staging involves evaluating cellular features and the tumor's spatial distribution and extent, often across multiple slides or regions. Some studies have applied deep learning to aspects of ovarian cancer pathology, such as subtype classification or survival prediction from images [12]. Still, further development requires a comprehensive automated staging system directly from brightfield thin tissue pictures for the FIGO stages. Challenges include the heterogeneity of ovarian cancer, the variability in morphological appearance across stages, and the need for models to handle large whole-slide images [13].

This study aimed to develop and validate a deep learning model that automatically predicts ovarian cancer FIGO surgical stage (mapped to 5 numerical classes: 0-4 for control and Stages I-IV) directly from the transmission microscopy images. We utilized a transfer learning approach with a ResNet-101 model and employed advanced training techniques to build a robust classifier, including data augmentation and hyperparameter optimization via a genetic algorithm. Our findings demonstrate that deep learning can achieve high accuracy in discriminating between different stages of ovarian cancer based on image morphology alone. The principal conclusion is that an AI-based system holds significant potential as a valuable tool to enhance the accuracy, consistency, and efficiency of ovarian cancer staging in pathology workflows.

## 2. Results and Discussions

This section details the deep learning model's performance for automated ovarian cancer stage prediction on the test dataset. The results are interpreted in the context of the model's ability to capture stage-defining morphological features and discussed in terms of existing work and clinical implications for ovarian cancer pathology using their brightfield images.

**2.1. Model Training Performance**

The training process involved optimizing the ResNet-101 model on the ovarian histopathology training set using thin tissue brightfield transmission microscopy images. Figure 1 displays the learning curves, showing the model's performance on the training and validation datasets.
As illustrated in Fig. 1, the training loss consistently decreased throughout the training process, indicating effective learning from the training data. Concurrently, the validation loss initially decreased before stabilizing, and the validation accuracy improved, reaching a peak value.

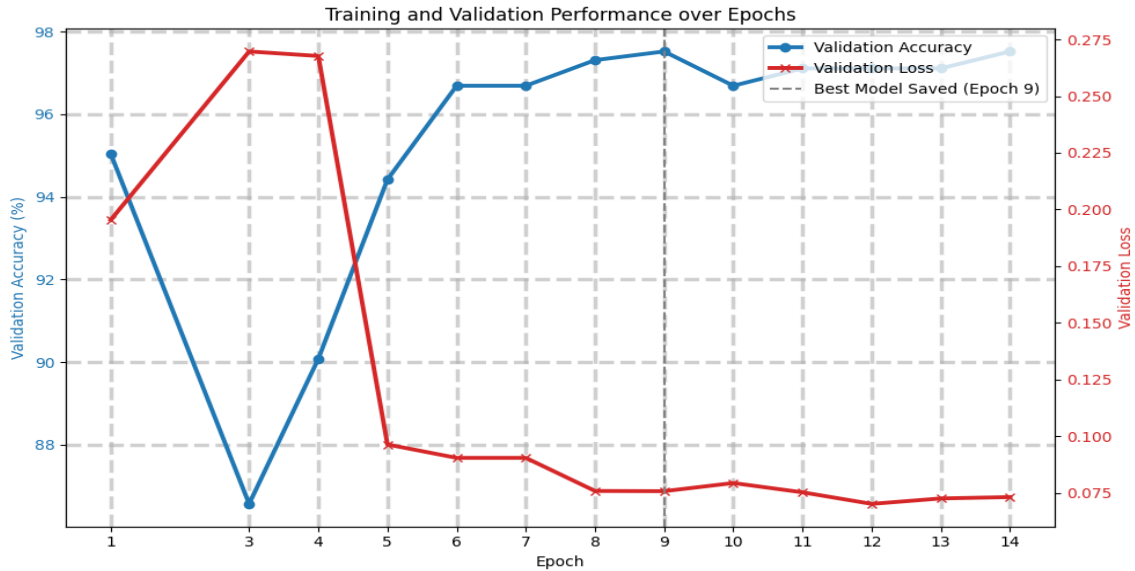

**Figure 1. Training and Validation Loss/Accuracy Curves.** This figure presents the model's validation performance across 14 training epochs. The blue curve represents validation accuracy, while the red indicates validation loss. A noticeable performance improvement is observed after epoch 4, where the validation loss drops significantly and accuracy increases steadily. The model achieved its best performance at epoch 9, as indicated by the dashed vertical line, corresponding to the highest validation accuracy. These trends suggest effective learning and successful generalization, with minimal signs of overfitting throughout the training process.

The early stopping mechanism was configured to monitor validation accuracy and halt training at Epoch '14' after '9' epochs without significant improvement, preventing overfitting and identifying the optimal model state. This convergence behavior on the validation set suggests the model successfully learned generalizable features pertinent to ovarian cancer staging without merely memorizing the training examples.

## 2.2. Test Set Performance Metrics

The final performance of the trained model was assessed on a completely independent test set consisting of 126 ovarian bright-field images. The model achieved a high overall classification accuracy of 97.62%, corresponding to 123 correctly classified samples out of the 126 test samples. Table 1 provides a comprehensive classification report detailing each stage's precision, recall, F1-score (Class 0 through Class 1V), and overall average metrics.

| | Precision | Recall | F1-Score | Support |
|---|---|---|---|---|
| Class 0 | 1.0 | 1.0 | 1.0 | 27 |
| Class 1 | 0.97 | 0.98 | 0.98 | 61 |
| Class 2 | 0.91 | 0.91 | 0.91 | 11 |
| Class 3 | 1.0 | 0.95 | 0.97 | 19 |
| Class 4 | 1.0 | 1.0 | 1.0 | 8 |
| accuracy | | | 0.98 | 126 |
| macro avg | 0.98 | 0.97 | 0.97 | 126 |
| weighted avg | 0.98 | 0.98 | 0.98 | 126 |

**Table 1.** The table summarizes the model's performance metrics across all five classes regarding precision, recall, and F1-score. Class 0 and Class 4 achieved perfect scores (1.0) across all metrics, indicating flawless classification. Class 1 and Class 3 were also predicted with high accuracy, with F1-scores of 0.98 and 0.97, respectively. Class 2 had the lowest performance but achieved solid values (0.91 for all three metrics). The overall model performance was strong, with an accuracy of 0.98. The macro and weighted averages confirm consistent and balanced performance across the dataset.

(Note: The script outputted '0.98' for accuracy, macro avg, and weighted avg, 97.62% rounded to two decimal places. It's standard practice to report the exact output of the metric calculation tool in the table.)

An overall accuracy of 97.62% on an independent test set represents a strong result for automated ovarian cancer staging from the bright-field histopathology images. This high accuracy suggests the model has learned to effectively discern the complex morphological features associated with different stages of ovarian cancer spread. Examination of the classification report (Table 1) reveals the model's performance across individual stages. The model demonstrated excellent performance on Class 0, Class 3, and Class 4, achieving high precision (1.00 for Classes 0, 3, 4) and recall (1.00 for Classes 0, 4; 0.95 for Class 3) for these stages, resulting in F1-scores of 1.00 or 0.97. Specifically, it achieved perfect precision (1.00) for Classes 0, 3, and 4, indicating that predictions of these stages were always correct. It also showed perfect or near-perfect recall (1.00 for Classes 0 and 4, 0.98 for Class 1), demonstrating its ability to identify most instances of these stages. The F1-scores, which balance precision and recall, demonstrate robust performance for Class 0 (1.00), Class 1 (0.98), Class 3 (0.97), and Class 4 (1.00). The slightly lower F1-score for Class 2 (0.91) suggests that distinguishing this stage might be marginally more challenging for the model than the others.

### 2.3. Confusion Matrix Analysis

A confusion matrix (Figure 2) was generated based on the predictions on the test set to understand the model's performance further and identify specific misclassification patterns.
Fig. 2 presents the confusion matrix, showing the actual versus predicted stages distribution. Out of 126 test samples, the model correctly classified 123. The misclassifications were minimal and primarily occurred between adjacent stages. Specifically, 1 sample from the actual Class 1 was misclassified as

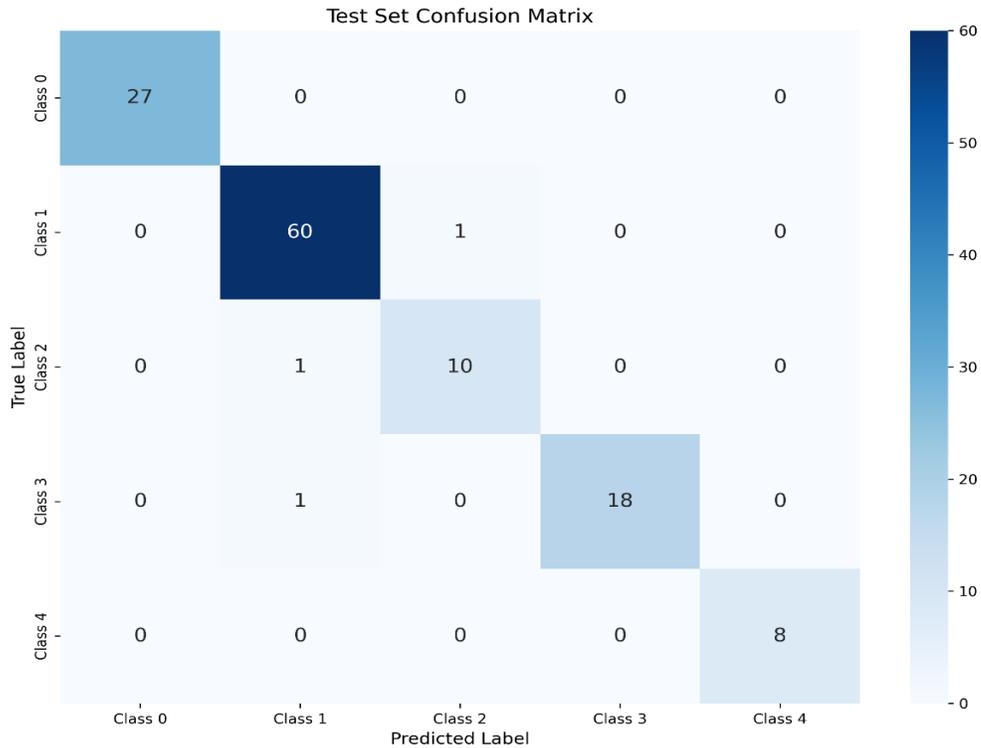

**Figure 2: Test Set Confusion Matrix.** This matrix illustrates the performance of the trained model in classifying test samples into five distinct classes (Class 0 to Class 4). The diagonal elements represent the number of correctly classified instances for each class, while the off-diagonal elements indicate misclassifications.

Class 2, 1 sample from the actual Class 2 was misclassified as Class 1, and 1 sample from the actual Class 3 was misclassified as Class 1. There were no misclassifications involving Class 0 or Class 4, for which the model achieved perfect precision and recall. The clustering of errors between adjacent stages (Class 1, 2, and 3) is consistent with the subtle morphological differences that can exist between these stages, sometimes presenting challenges even for human pathologists. The ability to correctly classify all samples of the extreme stages (0 and 4) is a powerful indicator of the model's learned features. A visual representation of this matrix is provided in Figure 2.

The misclassifications between adjacent stages might stem from inherent morphological similarities or variations within a stage [14]. For instance, Stage II involves pelvic extension, while Stage III consists of spreading to the upper abdomen or lymph nodes; distinguishing these might rely on evaluating features indicative of localized invasion versus widespread peritoneal involvement, which can be subtle on image patches alone. This mirrors challenges sometimes encountered in manual pathology, where distinguishing borderline cases or interpreting limited biopsy samples can be difficult [15]. The model's performance aligns with the task's complexity, suggesting it has learned relevant stage-associated morphological cues, even if the boundaries between stages remain the most challenging to delineate automatically.

## 2.4. Potential Clinical Implications

An automated system capable of predicting ovarian cancer stage with high accuracy has significant potential clinical implications. It could be a valuable tool in digital pathology workflows using brightfield microscopy images, potentially providing an initial automated staging suggestion for every case. This could help prioritize cases for pathologist review, flag potential discrepancies, and contribute to greater consistency in reporting, especially in high-throughput settings or for less experienced pathologists. By automating the initial assessment, it could also contribute to faster, more accurate diagnostic turnaround times, which is crucial for prompt treatment initiation in aggressive malignancies like ovarian cancer.

It is essential to emphasize that this AI tool is envisioned as an aid to the pathologist, not a replacement. The pathologist integrates imaging findings with clinical history, macroscopic evaluation, intraoperative findings, and potentially other ancillary tests. An automated system could serve as an objective enhancer to this process. Future clinical implementation would require prospective validation, user interface design, and regulatory approval. Furthermore, while the model performs well, understanding the specific visual cues it uses (model interpretability) would increase pathologists' trust and facilitate diagnostic refinement.

# 3. Methodology and Instrumentation

This section describes the experimental setup for large-volume bright-field imaging, including the dataset characteristics, data preprocessing, model architecture, training configuration, and evaluation procedures used to develop and test the automated ovarian cancer staging framework.

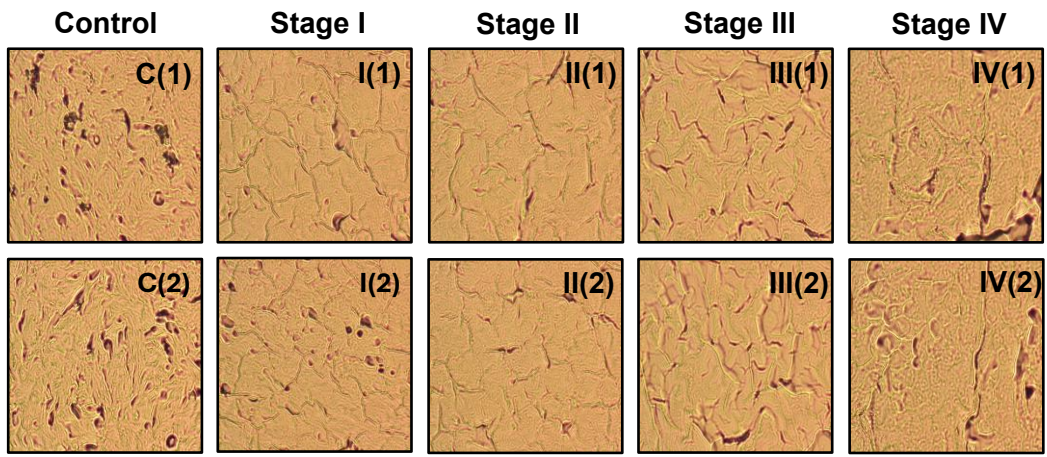

**Figure 3.** Representative transmission optical brightfield images from five different classes: Control C (1,2), Stage I (1,2), Stage II(1,2), Stage III(1,2), and Stage IV (1,2). Two representative images are shown for each class, illustrating the differences in tissue structure across the control and ovarian cancer progression stages.

We obtained ovarian tissue samples in the form of tissue microarray (TMA) slides from Biomax.us. Each slide contained 24 thin tissue cores, measuring 1.5 mm in diameter and 5 microns in thickness. We conducted imaging using an Olympus BX61 microscope operated in transmission mode and equipped with a 40X objective lens. To capture the images, we used an AmScope digital camera. The scanning process was fully automated using the OptiScan software from Prior Test, which enabled precise synchronization between the microscope stage movement and image acquisition. Details are described in Ref. [16]. Using this setup, we achieved a scanning rate of approximately 100 images every 6 minutes, allowing for an efficient and consistent collection of brightfield images as shown in Fig. 3.

### 3.1. Dataset Acquisition and Description

The dataset utilized in this study comprised images sourced from two distinct collections to create the training/validation partition and an independent test set. The primary collection, used for model training and validation, consisted of 2418 brightfield images of ovarian thin tissue samples. These images were split into training and validation sets. Each image was associated with a corresponding FIGO surgical stage, mapped to five numerical classes: 0, 1, 2, 3, and 4 corresponding to control, Stage I, Stage II, Stage III and Stage IV. An independent test set, comprising 126 images, was obtained from BX61. This independent set was used solely for the final, unbiased evaluation of the trained model. The class distribution within the independent test set was: Stage 0 (27), Stage I (61), Stage II (11), Stage III (19), and Stage IV (8). The combined distribution of images across all three partitions and five stages is detailed in Appendix A (Table A1).

### 3.2. Image Preprocessing and Data Augmentation

All histopathological images underwent essential preprocessing steps before being fed into the deep learning model. Images were uniformly resized to a standard input dimension of (224, 224) pixels.

A comprehensive suite of data augmentation techniques was applied dynamically during training for the training dataset. These augmentations were crucial for increasing the adequate size of the training dataset, enhancing the model's robustness, and improving its ability to generalize to variations present in real-world histopathology slides. The specific augmentations applied were:

- Random Horizontal Flip (with a probability of 0.5)
- Random Vertical Flip (with a probability of 0.5)
- Random Rotation (applied within a range of ±15 degrees)
- Color Jitter (introducing random variations in brightness, contrast, and saturation, each within a factor of 0.1)
- Random Affine Transformations (including random translation, scaling between 0.8 and 1.2, and shear up to 10 degrees)

These transformations simulate common variabilities encountered in tissue preparation and imaging, such as different orientations, slight morphological distortions, and variations in staining intensity. Following these geometric and color manipulations (applied stochastically per image in the training set), the images were converted to PyTorch tensors using transforms to tensor () and normalized.

The images underwent only the resizing and tensor conversion steps for the validation and independent test sets, followed by normalization. Normalization was performed using the standard mean ([0.485, 0.456, 0.406]) and standard deviation ([0.229, 0.224, 0.225]) values derived from the ImageNet dataset, consistent with the pre-training of the base ResNet-101 model. Data augmentation was explicitly not applied to the validation or test data to ensure unbiased evaluation of the model's performance on fixed, unseen samples.

## 3.3. Model Architecture

The chosen architecture for our automated ovarian cancer staging task was a ResNet-101 [22] convolutional neural network. ResNet architectures are well-regarded for their ability to train deep networks effectively, primarily due to the inclusion of skip connections that help mitigate issues like vanishing gradients during backpropagation. This makes them particularly effective for learning complex hierarchical features necessary for intricate image analysis tasks like histopathology interpretation.

We initialized the ResNet-101 model with weights pre-trained on the large-scale ImageNet dataset. This transfer learning approach leverages the powerful generic visual feature extraction capabilities learned from natural images. It adapts them for the specific domain of ovarian cancer histopathology, which is often beneficial when medical datasets are smaller than benchmark datasets like ImageNet. The original classification layer of the ResNet-101 was removed and replaced with a custom nn.Sequential module tailored explicitly for our 5-class ovarian cancer staging task (mapping to stages 0 through IV):

A dropout layer was strategically included immediately before the final linear layer for regularization. This helps prevent the model from overfitting to the specific nuances and noise of the training data by randomly setting a fraction of input units to zero during training. The dropout probability, denoted as BEST_DROPOUT_RATE, was determined as the optimal value using the genetic algorithm by the hyperparameter optimization process. This value was found to be 0.30315513401405514. The final layer

```
model.fc = nn.Sequential(
    nn.Dropout(BEST_DROPOUT_RATE),
    nn.Linear(model.fc.in_features, NUM_CLASSES)
)
```

transforms the high-level feature vector extracted by the ResNet backbone into a 5-dimensional output vector. Each element in this vector corresponds to the predicted score for one of the five cancer stages (0-IV), which is then passed through a Softmax function to obtain probability distributions over the classes.

## 3.4. Training Setup

The final training of the model was performed using the Adam optimizer. The initial learning rate was set to the optimal value determined by the genetic algorithm (Section 3.5), precisely $2.00 \times 10^{-4}$. Weight decay, also optimized by the GA, was set to $3.72 \times 10^{-5}$ to help regularize the model by penalizing large weights.

To effectively handle the potentially uneven distribution of samples across the five cancer stages in the training dataset, we employed a weighted random sampler (WeightedRandomSampler) and a weighted loss function. Weights for the sampler were computed for each sample as inversely proportional to the frequency of its corresponding class in the training data. This ensured that instances from minority classes were sampled more frequently during batch creation. The loss function used was the Cross-Entropy Loss (nn.CrossEntropyLoss), which was also weighted using class weights computed inversely proportional to the class frequencies. This assigns higher penalties for misclassifying samples from underrepresented classes, encouraging the model to learn their features more effectively.

Training proceeded for a maximum of 20 epochs. A learning rate scheduler, specifically ReduceLROnPlateau, was utilized to adapt the learning rate during training. The learning rate was reduced by a factor of 0.1 if the validation accuracy did not show improvement for two consecutive epochs. Early stopping was critical to the training process, monitoring the validation accuracy. Training was terminated if the validation accuracy did not improve for five successive epochs, and the model weights corresponding to the epoch with the highest validation accuracy achieved up to that point were loaded. Training was performed at a dedicated GPU.

## 3.5. Hyperparameter Optimization

To determine the most effective configuration for key training parameters, a genetic algorithm (GA) was implemented to optimize the learning rate for the Adam optimizer, the dropout rate in the final classification layer, and the weight decay parameter. The objective of the GA was to identify the combination of these hyperparameters that yielded the highest validation accuracy for the model.

The GA was configured with a population size of 5 individuals. Everyone represented a unique combination of the three hyperparameters, sampled initially from their predefined search spaces: learning rate within [1e-5, 1e-3], dropout rate within [0.3, 0.7], and weight decay within [1e-6, 1e-4]. The population evolved over three generations.

Everyone's fitness was evaluated by training the model using its specific set of hyperparameters for a reduced number of epochs (5 epochs) on the training data and measuring the resulting validation accuracy. Based on these fitness scores, a selection process was applied. We used tournament selection with a size of 3 to select parent individuals who would contribute to the next generation.

New individuals (offspring) were generated through blend crossover, combining the hyperparameter values of two selected parents. Genetic diversity was maintained through Gaussian mutation, where a small amount of Gaussian noise was added to the hyperparameters of offspring with a probability determined by the mutation rate, set at 0.1. This evaluation, selection, crossover, and mutation process was repeated for the specified generations.

The optimal hyperparameters identified by the genetic algorithm, which demonstrated the highest validation accuracy during the GA search phase, were:

- Learning rate: 0.00020027035366007752
- Dropout rate: 0.30315513401405514
- Weight decay: 3.724157568008995e-05

These optimized values were subsequently used for the final training run of the model as described in Section 3.4.

The final trained model's performance on the independent test set was quantitatively assessed using standard classification metrics. These included overall accuracy, per-class precision, recall (sensitivity), and F1-score. A confusion matrix was also generated to visualize the pattern of correct and incorrect predictions across the five ovarian cancer stages. Calculations were performed using the sklearn.metrics library in Python.

## 4. Conclusions and Discussion

In this study, we successfully developed and evaluated a deep learning framework for automated ovarian cancer staging directly from thin tissue brightfield images. The model, built upon a fine-tuned ResNet-101 architecture and optimized using a genetic algorithm for key hyperparameters including learning rate, dropout rate, and weight decay, achieved a high overall accuracy of 97.62% on an independent test set comprising five cancer stages (0-IV). This performance powerfully demonstrates the capability of artificial intelligence to effectively analyze the complex morphological patterns present in ovarian histopathology, indicative of cancer progression and the extent of spread.

The high classification accuracy obtained is particularly promising for clinical application. By providing a rapid and objective initial stage assessment, this automated system has the potential to serve as a valuable aid in the diagnostic process for ovarian cancer. It could contribute to standardizing reporting, potentially reducing the well-documented inter-observer variability encountered in manual pathological assessment, and improve overall efficiency in pathology laboratories, especially those handling high numbers of cases. For a malignancy where timely and accurate staging is paramount for guiding treatment decisions and predicting patient prognosis, such a tool could have a meaningful impact on patient care. The model's training on a dedicated GPU facilitated the computational intensity for exploring hyperparameters and training the final high-performance model.

Analysis of the detailed performance metrics further supported the model's capabilities. The classification report (Table 1) showed high precision, recall, and F1-scores across most stages, with excellent performance noted for Stage 0 (Benign/Normal) and Stage 4, achieving F1-scores of 1.00. The confusion matrix (Table 2 and Figure 2) provided deeper insight, revealing that most rare misclassifications (only 3 out of 126 test samples) occurred between adjacent stages, specifically confusing stages from Classes 1, 2, and 3. This finding is significant as it aligns with the clinical reality that distinguishing between adjacent FIGO stages of ovarian cancer can sometimes involve subtle morphological cues or depend heavily on assessing the precise extent of spread rather than just cellular features, which can be challenging even for experienced pathologists [5, 6]. The model's ability to correctly classify samples from the extreme stages (0 and 4) with perfect accuracy is a powerful indicator of its learned feature representations.

Despite these strong results, several limitations and areas for future work must be considered. Validation on larger, more diverse multi-institutional datasets encompassing a wider range of patient demographics, different histological subtypes of ovarian cancer, and varying tissue processing and imaging protocols is essential to ensure the model's generalizability and robustness in real-world clinical settings [17]. Furthermore, while the model performs well on image patches, integrating information across entire Whole Slide Images (WSIs) or combining image analysis with other relevant clinical or pathological data (e.g., gross pathology descriptions, tumor markers, radiologic findings) could potentially improve performance, particularly for distinguishing between morphologically similar adjacent stages [18]. Integrating this AI tool into existing digital pathology workflows would require developing user-friendly interfaces and ensuring robust pathologist review and override mechanisms. Finally, exploring methods to enhance the interpretability of the model's predictions would provide valuable insights for pathologists and build necessary trust in AI-assisted diagnostics.

Despite these challenges and future avenues for exploration, the results presented here prove that deep learning holds significant promise for automating crucial aspects of ovarian cancer digital pathology, specifically surgical staging. As digital pathology workflows continue to expand globally, AI tools like the one developed in this study are poised to play an increasingly important role in enhancing diagnostic accuracy, consistency, and efficiency, ultimately contributing to improved patient care and outcomes for women diagnosed with ovarian cancer [19].

**Appendix A**

This appendix provides supplementary details regarding the distribution of histopathological images across the different cancer stages within the training, validation, and independent test sets used in this study. This information is essential for the reproducibility of the experimental setup. The dataset consists of five classes, corresponding to cancer stages 0 (Benign/Normal), and Stages: I, II, III, and IV.

The original dataset was partitioned into training and validation sets using a stratified 80/20 split. The independent test set was sourced separately. As shown in Table A1, stratification ensured that the class distribution was maintained proportionally across the training and validation subsets relative to their source (train.csv). The test set, comprising 126 images, had a detailed distribution. This stratified splitting approach and techniques like weighted random sampling during training helped address potential class imbalance and ensured robust evaluation.

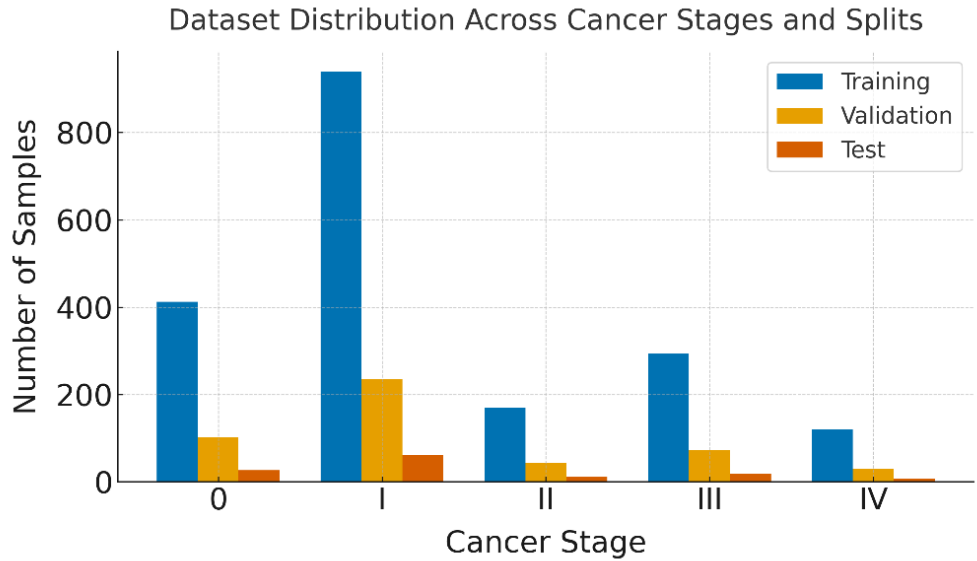

**Figure A**1. This figure presents the number of brightfield images for each cancer stage (0 to IV) distributed across the training, validation, and test sets. Each bar represents the count of images per stage in each split, providing a clear overview of the dataset composition used for model development and evaluation.

## Appendix B

This appendix provides additional details on the genetic algorithm's (GA) performance while searching for optimal hyperparameters for the deep learning model. Beyond reporting the final best parameters in the main text (Section 3.5), illustrating the search trajectory offers insights into the optimization process.

Figure B1 illustrates the evolution of the validation accuracy (fitness) for the best and average individuals in the population across the three generations of the genetic algorithm.

Figure B1 demonstrates the progress of the genetic algorithm in optimizing the hyperparameters for improved model performance on the validation set. The 'Best Fitness' line tracks the highest validation accuracy achieved by any individual within a given generation's evaluation phase. The 'Average Fitness' line shows the mean validation accuracy across all five individuals in the population for each generation. The increase in average fitness over generations indicates that the algorithm effectively guided the search towards more promising regions of the hyperparameter space. The best fitness fluctuated across generations, reflecting the nature of stochastic optimization algorithms, with the highest fitness in Generation 3 being the basis for selecting the hyperparameters for the final training run (Section 3.4).

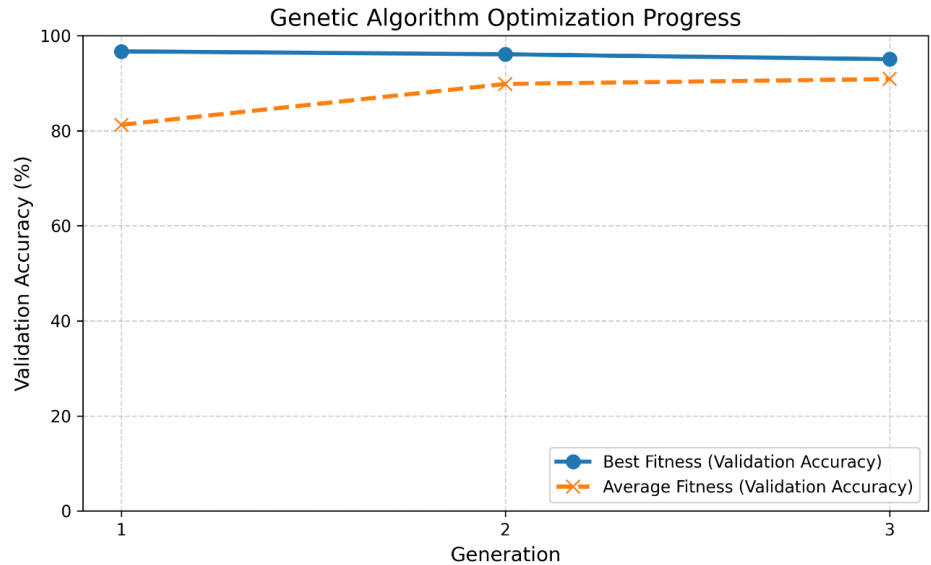

**Figure B1** illustrates the genetic algorithm's optimization progress over three generations. The blue curve represents the best validation accuracy (best fitness) achieved in each generation, while the orange dashed curve shows the population's average validation accuracy (average fitness). The plot highlights how the algorithm improves the average performance over generations while maintaining a consistently high best fitness value.

**Authors Contributions**:
PP conceived the idea and project; MA performed the large-scale brightfield imaging; AM performed the AI analyses, and MA helped him. AM wrote the first draft of the paper, MA and PP edited it, and all authors contributed to the final version.

**Acknowledgments**
We acknowledge NIH, MSU, and ORED(MSU) for funding the project.

**Funding**
This work was partially supported by the National Institutes of Health under grant R21 CA260147 and ORED, Mississippi State.

**Data Availability Statement**
Data are available from the corresponding author upon reasonable request.

**Conflicts of Interest**

The authors declare no conflicts of interest.